\documentclass[a4paper]{article}

\usepackage{INTERSPEECH2018}

\usepackage{setspace}       
\usepackage{array}          
\usepackage[utf8]{inputenc} 
\usepackage[T1]{fontenc}    
\usepackage{hyperref}       
\usepackage{url}            
\usepackage{booktabs}       
\usepackage{amsfonts}       
\usepackage{nicefrac}       
\usepackage{microtype}      
\usepackage{xspace}         
\usepackage{amsmath}        
\usepackage{graphicx}       

\newcommand{\eg}{\textit{e.g.,\ }}
\newcommand{\etc}{\textit{etc.\@}\xspace}

\newcommand{\vs}{\textit{vs.\ }}

\title{Play Duration based User-Entity Affinity Modeling in Spoken Dialog System}
\name{Bo Xiao$^1$, Nicholas Monath$^2$, Shankar Ananthakrishnan$^1$, Abishek Ravi$^1$}
\address{
  $^1$Amazon.com, Cambridge, MA, U.S.A.\\
  $^2$CICS, University of Massachusetts Amherst, Amherst, MA, U.S.A.}
\email{$^1$\{boxiao, sanantha, abishekr\}@amazon.com, $^2$nmonath@cs.umass.edu}

\begin{document}

\maketitle
\begin{abstract}
Multimedia streaming services over spoken dialog systems have become ubiquitous.
User-entity affinity modeling is critical for the system to understand and disambiguate user intents and personalize user experiences.
However, fully voice-based interaction demands quantification of novel behavioral cues to determine user affinities.
In this work, we propose using play duration cues to learn a matrix factorization based collaborative filtering model.
We first binarize play durations to obtain implicit positive and negative affinity labels.
The Bayesian Personalized Ranking objective and learning algorithm are employed in our low-rank matrix factorization approach. 
To cope with uncertainties in the implicit affinity labels, we propose to apply a weighting function that emphasizes the importance of high confidence samples.
Based on a large-scale database of Alexa music service records, we evaluate the affinity models by computing Spearman correlation between play durations and predicted affinities.
Comparing different data utilizations and weighting functions, we find that employing both positive and negative affinity samples with a convex weighting function yields the best performance.
Further analysis demonstrates the model's effectiveness on individual entity level and provides insights on the temporal dynamics of observed affinities.
\end{abstract}

\noindent\textbf{Index Terms}: collaborative filtering, Bayesian personalized ranking, voice-controlled device, implicit affinity

\section{Introduction}
\label{sec:intro}
Fully voice-controlled consumer products such as Amazon Echo have connected millions of users to their desired multimedia services through the spoken dialog system.
Challenges remain when the system attempts to resolve ambiguities.
One type of ambiguity is inherent.
The same mention may stand for multiple valid entities.
For example, ``\textit{play hello}'' may refer to the song either by Adele or by Lionel Richie.
Without screen and keypad, it would be a poor user experience to read out the list of candidate entities and have the user select by voice.
The other type of ambiguity comes from errors in Spoken Language Understanding (SLU), incurred by the Automatic Speech Recognition (voice to text) and Natural Language Understanding (semantic interpretation) components.
Although multimedia entity retrieval engines are designed to be robust to minor distortion in the query, it is still vulnerable when an erroneous entity accidentally matches a noisy query.
To address these challenges, we investigate user-entity affinity models for an additional source of meta-information.
Estimating a user's preference to each entity may facilitate personalized entity resolution, \eg mapping the user's mention of ``\textit{hello}'' to the song from their favored artist.
Moreover, a user's mention of ``\textit{hello}'' may be decoded as ``hello'' and ``halo'', resulting in multiple candidate entities to play.
The affinity information may aid ranking SLU interpretations containing multiple hypotheses of user mentions and intents that lead to distinct entities. 

In the literature, extensive studies have been conducted on Collaborative Filtering (CF) based large scale user-entity affinity modeling \cite{sarwar2001item}.
CF models leverage the similarities of users and entities to infer unobserved affinities, given sparse observations of user-entity interactions.
Early methods represented each user as a vector of entities that the user had interacted with, and vice versa \cite{koren2008factorization, linden2003amazon}.
Unobserved affinity to an entity may be predicted by a weighted sum of the user's affinities to the neighbors of the entity that the user has interacted with. 
In addition, Matrix Factorization (MF) methods have been found to outperform the aforementioned neighborhood-based methods \cite{bennett2007netflix, koren2009matrix}.
MF models the users and entities in a common, low dimensional vector space such that the inner product of these vectors (\textit{a.k.a. embeddings}) approximates the user-entity affinity \cite{koren2008factorization}.

Despite the generalizability of CF, applying it to spoken dialog systems is not straightforward.
Users interact with voice interface of the system and express affinities in a fundamentally different way than typical Web based applications.
In the past, CF models have employed various explicit or implicit behavioral cues \cite{hu2008collaborative}.
For example, in online shopping or multimedia streaming applications, purchasing behavior and user ratings are explicit signals \cite{linden2003amazon, zhou2008large}.
In online Web search, clicking and browsing behaviors such as frequency of click and dwell time on the page, \etc are extracted as implicit cues \cite{agichtein2006improving, shen2012personalized, cai2017behavior}. 
In the scenario of voice-based search, previous studies have mainly focused on improving robustness towards noisy queries \cite{rao2017talking, levitan2014detecting, shokouhi2016did}.
However, there has not been much work on affinity cue quantification.
Since \textit{pay-per-view} transactions are typically not adopted in voice-based services, explicit affinity cues are not generally available, or tend to be very sparse.

We refer to the multimedia streaming event triggered by a user request as a \textit{playback}.
The playback may be completed by finishing the content or terminated early by the user.
The total time of such streaming event is defined as the \textit{play duration}.
We hypothesize that play duration may be an implicit behavioral cue for affinity, which is analogous to, but substantially different from, dwell time in Web browsing.
In Web browsing, users can proactively choose the page to view by clicking or tapping, so that even when the dwell time is short, it may still indicate some positive albeit low relevance.
If nothing is relevant, the user may not view any page, and no dwell time would be recorded.
In spoken dialog systems, users only receive resolutions passively; a playback does not necessarily indicate any positive affinity.
Terminated playback may attribute to an acceptable but not favored entity, \eg requesting the remix version but getting the original version of a song; or a wrong SLU interpretation and totally irrelevant entity, \eg playing the song \textit{Halo} when the mention is \textit{hello}.
We include the former as a part of \textit{positive} affinity, and define the latter as \textit{negative} affinity.
Empirically, negative affinity should cause much shorter play duration as users tend to immediately stop an irrelevant playback.
Capturing negative affinity is critical to improving user experience.
This motivates us to consider binarization of play durations to extract positive and negative affinities.

Play duration is implicit in representing affinity because there are external factors infeasible to capture.
Users may have various reasons to stop the playback early or leave it unattended.
Heuristically, confidence of positive affinity is higher when the play duration is much longer, and confidence of negative affinity is higher when the play duration is shorter.
This inspires us to add a weight \cite{hu2008collaborative} to each observation according to the confidence of the positive/negative label.

Following these ideas, in this work we investigate MF based CF modeling of user-entity affinity, using play duration cues.
As a case study, we focus on Alexa music services, though the modeling approach is agnostic to application domain or entity type.
In the rest of the paper, we explain the modeling approaches in Section \ref{sec:mf}; describe the data set, experiment setting, and evaluation results in Section \ref{sec:exp}; and conclude the paper with future directions in Section \ref{sec:con}.

\section{Matrix factorization model}
\label{sec:mf}
Let the set of users and entities be $U$ and $E$, respectively.
We denote user-entity affinities in matrix $A\,:\,|U|\times|E|$.
A cell $A_{u,e}$ in $A$ denotes affinity between $u$ and $e$.
Our goal is to derive $\mathbf{f}_{u}\,:\,1\times K$ for each $u\in U$, and $\mathbf{g}_{e}\,:\,K\times 1$ for each $e\in E$, such that their inner products approximate $A$.
We define the affinity score $\hat{a}_{u,e} = \mathbf{f}_{u}\cdot\mathbf{g}_{e}$ \footnote{Constant biases for each $u$ and $e$ are not included as the goal is not in rating estimation or recommendation.}.
The size of $\{\mathbf{f}_{u}\}\cup\{\mathbf{g}_{e}\}$ dominates the model's complexity in $O(|U|+|E|)$.

Directly representing affinities in the form of $A_{u,e}$ has some challenges.
First, distinct $(u,e)$ tuples may have dramatically different amount of observations and variances of $A_{u,e}$ estimation, so that the confidence of $A_{u,e}$ varies.
Moreover, holding $A_{u,e}$ values requires $O(|U|m)$ spatial complexity, where $m$ is the average count of entities played by a user.
This is dramatically larger than the model size. And so,
we label each interaction separately and process the observations sequentially.

\subsection{Implicit positive and negative affinity}
\label{sec:mf:bin}
We discretize observations into implicit positive/negative labels by an entity type specific threshold $T$, which is based on a certain percentile (omitted) of play durations for that entity type.
For example, for music domain entities, $T=30$ sec for song track, and $T=180$ sec for radio station and album.
Let $R$ denote the entire observation set.
For a request $r\in R$, we denote its play duration as $t_{r} \ge 0$, and the observed affinity label $a_{r}$ as in (\ref{equ:aff}).

\begin{equation}
\label{equ:aff}
  a_{r}=
  \begin{cases}
    1, & \text{if}\ t_{r} \ge T\, ; \\
    -1, & \text{otherwise}
  \end{cases}
\end{equation}

\subsection{BPR training samples}
\label{sec:mf:bpr}
We optimize the Bayesian Personalized Ranking (BPR) \cite{rendle2009bpr} objective to fit our MF model.
BPR minimizes pair-wise ranking errors of affinities, rather than point-wise reconstruction errors of $A$, resulting in better prevention of overfitting.
Typically, one BPR training example is composed by a given user and a pair of entities with opposite affinity labels.
In the past, BPR has been widely applied in learning MF models \cite{yin2017sptf, riedel2013relation, he2016vbpr}.
Although there are more advanced CF methods \cite{wu2017recurrent, sedhain2015autorec, liang2016factorization}, in this work we adopt this simple model to allow a clear experimental analysis of play duration cues in spoken dialog systems.

For our task, directly pairing positive and negative samples in $R$ for each user may lead to over-sampling the less frequent negative samples and  overfitting on those.
Also, it requires $O(|U|m)$ spatial complexity.
In this work, we employ a simple approach --- sampling entities randomly from $E$ as \textit{negative} peers for positive observations, and as \textit{positive} peers for negative observations.
Since the entity set is large, we expect a low chance to draw an entity that the user has ever interacted with.
We drop the sampled entity if it is identical to the entity in observation.
Let $R^{+}=\{r|r\in R, a_{r}=1\}$ and $R^{-}=\{r|r\in R, a_{r}=-1\}$ denote positive/negative observations, respectively.
We draw BPR training sample sets $S^{+}$ and $S^{-}$ as in (\ref{equ:trset}) and (\ref{equ:trneg}), where $n_{j}$ and $p_{j}$ are randomly sampled negative/positive entities, respectively.
Let $N$ be the count of training samples generated from each observation $r\in R$.

\begin{gather}
  S^{+} = \{(u_{r}, e_{r}, n_{j})\ |\ r\in R^{+},\ j=1,\cdots,N, \nonumber \\
  n_{j}\in E,\ n_{j}\neq e_{r}\} \label{equ:trset} \\
  S^{-} = \{(u_{r}, p_{j}, e_{r})\ |\ r\in R^{-},\ j=1,\cdots,N, \nonumber \\
  p_{j}\in E,\ p_{j}\neq e_{r}\} \label{equ:trneg}
\end{gather}

\subsection{Observation confidence weighting}
\label{sec:mf:wei}

\begin{figure}[htb]
  \centering
  \includegraphics[width=0.8\columnwidth]{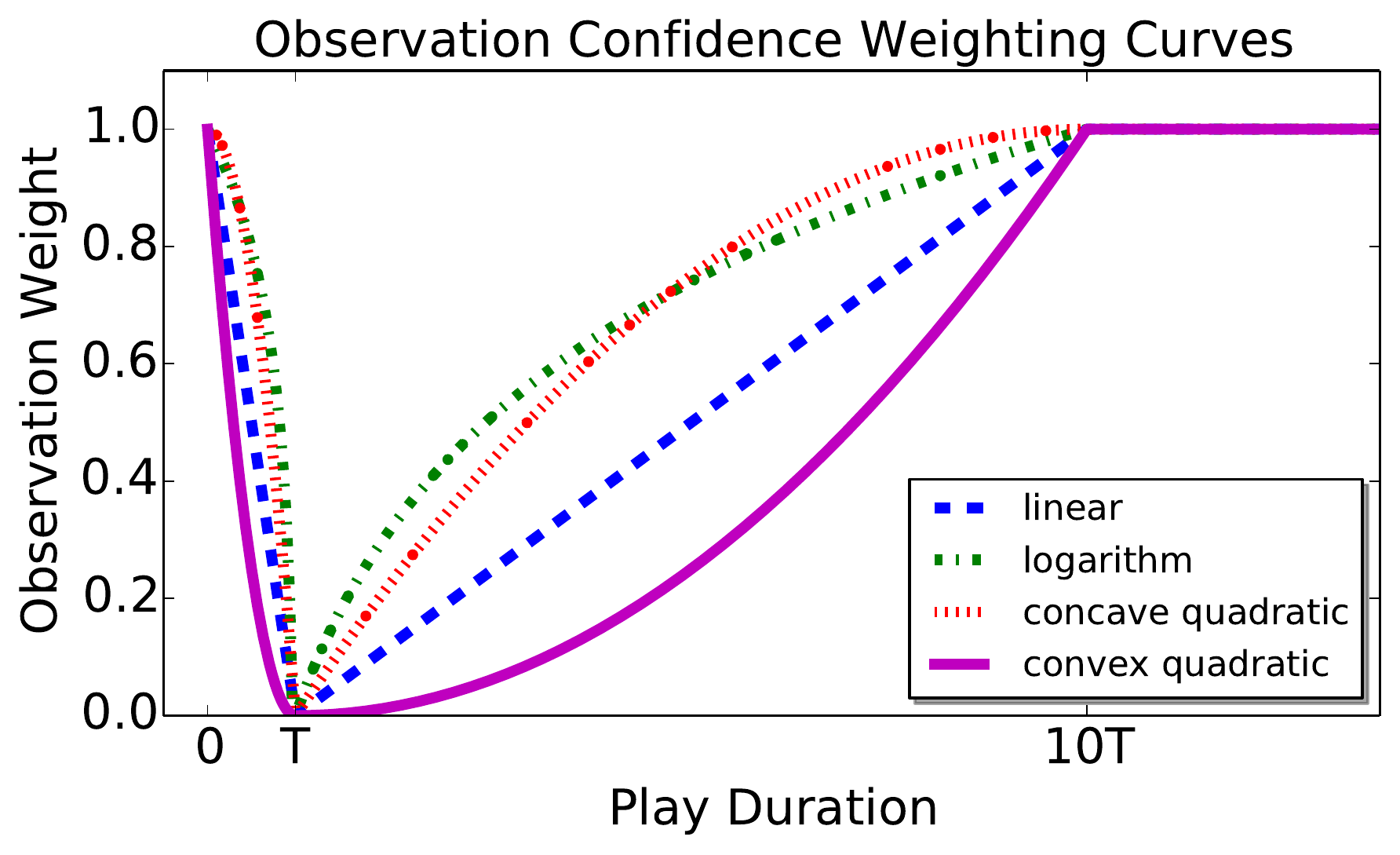}
  \caption{Observation confidence weighting curves. \label{fig:wei}}
\end{figure}

Binary labels do not capture fine grained notions of affinity, \eg a user stopping a playback at 1 sec \vs $T-1$ sec may indicate different levels of affinity.
We assign lower weights to samples closer to the threshold $T$.
Weight for each sample is derived from the play duration, through a function over the time axis.
Figure \ref{fig:wei} illustrates weighting functions in our investigation.
The weighting functions map $T$ to $0.0$, and reach $1.0$ at 0 sec and $10T$ separately (\eg 300 sec for song track, 1800 sec for album and station).
The weight, denoted $w(t_{r})$, remains a constant $1.0$ for $t_{r}>10T$.

\renewcommand{\arraystretch}{0.8}
\begin{table*}[tb]
  \caption{Illustration of Alexa music data set}
  \label{tab:data}
  \centering
  \begin{tabular}{cccccc}
    \toprule
    \textbf{Customer ID} & \textbf{Slot type} & \textbf{User mention} & \textbf{Resolution} & \textbf{Entity ID} & \textbf{Duration (sec)} \\
    \midrule
    User\_3 & SongName & hello & Hello & Entity\_5 & 10.0 \\
    \midrule
    User\_4 & StationName & n. p. r. & WBUR radio & Entity\_7 & 900.0 \\
    \midrule
    User\_9 & AlbumName & trolls & Trolls (OST) & Entity\_2 & 400.0 \\
    \bottomrule
  \end{tabular}
\end{table*}

\subsection{Objective function and optimization}
\label{sec:mf:opt}
Let $\sigma$ and $\mathbf{\Theta}$ denote the logistic function and the embeddings, respectively.
Let $\lambda_{\Theta}$ be the regularization parameter.
Let $\hat{x}_{uen}$ be the difference of affinity scores between positive entity $e$ \vs sampled negative entity $n$ for user $u$, as shown in (\ref{equ:bprest}).
Similarly let $\hat{x}_{upe}$ be that for negative observations as shown in (\ref{equ:bprneg}).
Let $w_{u,e}=w(t_{r})$ be the sample weight.
The objective function for BPR optimization is shown in (\ref{equ:bprobj}).

\begin{gather}
  \hat{x}_{uen}(\mathbf{\Theta})=\hat{a}_{u,e}(\mathbf{\Theta}) - \hat{a}_{u,n}(\mathbf{\Theta}) = \mathbf{f}_{u}\cdot(\mathbf{g}_{e} - \mathbf{g}_{n}) \label{equ:bprest} \\
  \hat{x}_{upe}(\mathbf{\Theta})=\hat{a}_{u,p}(\mathbf{\Theta}) - \hat{a}_{u,e}(\mathbf{\Theta}) = \mathbf{f}_{u}\cdot(\mathbf{g}_{p} - \mathbf{g}_{e}) \label{equ:bprneg} \\
  \sum_{(u,e,n)\in S^{+}}{w_{u,e}\texttt{ln}\sigma(\hat{x}_{uen}) - \lambda_{\Theta}||\mathbf{\Theta}||^{2}}\quad + \nonumber \\
  \sum_{(u,p,e)\in S^{-}}{w_{u,e}\texttt{ln}\sigma(\hat{x}_{upe}) - \lambda_{\Theta}||\mathbf{\Theta}||^{2} } \label{equ:bprobj}
\end{gather}

To allow for online updates, we optimize the model with stochastic gradient descent.
In (\ref{equ:bprsgd}), we illustrate the formula for updating the embedding values for the case of $S^{+}$ \cite{rendle2009bpr}.
The formula for $S^{-}$ only differs by subscripts.
We employ AdaGrad regularized dual averaging algorithm \cite{duchi2011adaptive} for regularization and learning rate adaptation. 

\begin{eqnarray}
  \mathbf{f}_{u} & \leftarrow & \mathbf{f}_{u} + \eta\cdot w_{u,e} \cdot (\sigma(-\hat{x}_{uen})(\mathbf{g}_{e} - \mathbf{g}_{n}) - \lambda_{\Theta}\mathbf{f}_{u}) \nonumber \\
  \mathbf{g}_{e} & \leftarrow & \mathbf{g}_{e} + \eta\cdot w_{u,e} \cdot (\sigma(-\hat{x}_{uen})\mathbf{f}_{u} - \lambda_{\Theta}\mathbf{g}_{e}) \label{equ:bprsgd} \\
  \mathbf{g}_{n} & \leftarrow & \mathbf{g}_{n} + \eta\cdot w_{u,e} \cdot (-\sigma(-\hat{x}_{uen})\mathbf{f}_{u} - \lambda_{\Theta}\mathbf{g}_{n}) \nonumber
\end{eqnarray}

In testing, we compute affinity prediction $\overline{a}_{u,e}$ in (\ref{equ:inf}) as the cosine similarity between $\mathbf{f}_{u}$ and $\mathbf{g}_{e}$, which is bounded between $-1.0$ and $1.0$.

\begin{equation}
\label{equ:inf}
  \overline{a}_{u,e} = \frac{\mathbf{f}_{u}\cdot\mathbf{g}_{e}}{||\mathbf{f}_{u}||\ ||\mathbf{g}_{e}||}
\end{equation}

\renewcommand{\arraystretch}{0.9}
\begin{table*}[!t]
  \caption{Spearman correlation between play durations and predicted affinities (showing \textbf{absolute differences})}
  \label{tab:res}
  \centering
  \begin{tabular}{c|cc|cc|ccc|cc}
    \toprule
      & \textbf{Data} & \textbf{Weight} & $\rho$ & $\widetilde{\rho}$ & $\rho_{\mathtt{song}}$ & $\rho_{\mathtt{album}}$ & $\rho_{\mathtt{station}}$ & $\mathtt{AUC}_{T}$ & $\mathtt{AUC}_{5T}$ \\
    \midrule
    (a) & $R^{+}_{\mathtt{m3}}$ & uniform & -- & $-0.068$ & -- & -- & -- & -- & $-0.0428$ \\
    \midrule
    (b) & $R_{\mathtt{m3}}$ & uniform & $+0.054$ & $+0.029$ & $+0.105$ & $+0.089$ & $+0.085$ & $+0.0773$ & $+0.0136$ \\
    \midrule
    (c) & $R_{\mathtt{m3}}$ & log & $+0.071$ & $+0.057$ & $+0.138$ & $+0.107$ & $+0.106$ & $+0.0883$ & $+0.0321$ \\
    \midrule
    (d) & $R_{\mathtt{m3}}$ & concave quad. & $+0.076$ & $+0.059$ & $+0.142$ & $+0.106$ & $+0.105$ & $+0.0884$ & $+0.0342$ \\
    \midrule
    (e) & $R_{\mathtt{m3}}$ & linear & $+0.086$ & $+0.070$ & $+0.155$ & $+0.112$ & $+0.112$ & $+0.0903$ & $+0.0406$\\
    \midrule
    (f) & $R_{\mathtt{m3}}$ & convex quad. & $+0.105$ & $+0.090$ & $+0.182$ & $+0.118$ & $+0.118$ & $+0.0863$ & $+0.0561$ \\
    \midrule
    (g) & $R_{\mathtt{m1\sim m3}}$ & convex quad. & $+0.137$ & $+0.125$ & $+0.221$ & $+0.164$ & $+0.138$ & $+0.1079$ & $+0.0749$ \\
    \bottomrule
  \end{tabular}
\end{table*}

\section{Experiments}
\label{sec:exp}

\subsection{Data set}
\label{sec:exp:data}
In this work, we use a subset of Alexa music service logs over a three-month period, containing requests to playable entities such as song track, radio station, and album.
The observed interactions are from far more than 100K users, amounting to over 10M requests/playbacks, and 100K distinct entities.
Table \ref{tab:data} illustrates fields in the data set with a few representative examples.
Slot type denotes system interpreted entity type.

Considering complexity and robustness, we do not include Customer/Entity IDs with less than 5 occurrences in the model.
Instead, we replace these IDs with ``\texttt{<User\_UNK>}'' and ``\texttt{<Entity\_UNK>}'' labels.
Ratio of requests with \texttt{UNK} index is below 5\%.
We expect the \texttt{UNK} index to capture an averaged characteristic of low frequency user/entity.
Finally, for anonymization and computational efficiency, we replace all Customer/Entity IDs including \texttt{UNK} by integers and store them in a vocabulary.

\subsection{Evaluation method}
\label{sec:exp:eva}

We compute the following two metrics for evaluation.
The first is Spearman correlation $\rho$ between the predicted affinities and the play durations.
Note that Spearman correlation tolerates the absolute values and only computes correlation of ranks, suitable for our problem as play duration may not have a linear relation to the latent affinity.
As the lengths of song tracks are much shorter than those of stations and album, we normalize play durations by dividing the entity type specific thresholds $T$, and denote the correlation in this case as $\widetilde{\rho}$.
Entity type specific $\rho_{type}$ values are also evaluated.

Secondly, we simulate prediction of the affinity labels by the predicted affinity.
By varying the decision boundary on the cosine similarities, we can plot the ROC curve for \textit{True Positive Rate} --- true positive prediction out of actual positive samples \vs \textit{False Positive Rate} --- false positive prediction out of actual negative samples.
We then report the Area Under ROC Curve (AUC) metric.
We may also modify the thresholds in (\ref{equ:aff}), then recompute the AUC metric as the affinity labels are modified.
This allows evaluation of the model's discriminative power towards different ranges of play duration.

\subsection{Experiment setting}
\label{sec:exp:set}
We first only use the third month (m3) to gain insights in the experiment, than investigate if more data (m1 $\sim$ m3) increases the accuracy.
In a causal setting, we reserve the second last day in m3 for development, and reserve the last day for testing.
Implementation was based on FACTORIE library \cite{mccallum2009factorie}.
To further speed up experiments, we employed the Hogwild trainer for parallel computing \cite{recht2011hogwild}.
For unseen users or entities in the test set, we substitute their IDs with the \texttt{UNK} index; thus the affinity prediction is always computable.

We optimized hyper-parameters by grid-search on the dev-set, including dimension of embedding $K$, learning rate $\eta$, regularization $\lambda_{\Theta}$, negative sampling count $N$, training iterations $I$, \etc 
As a result, $K=50$, $\eta=0.2$, $\lambda_{\Theta}=0.005$, $N=3$, $I=5$ yielded the optimal performance.
The relatively low values of $K$ and $N$ are favorable for large scale application.
The superior result from a relatively small $N$ is probably because of less collision to entities in the same polarity of affinity in the random sampling.

\subsection{Overall results}
\label{sec:exp:res}

Table \ref{tab:res} tags settings from (a) to (g).
The first two columns denote data utilization and sample weighting method.
The remaining columns report $\rho$, $\widetilde{\rho}$, and AUC of predicting $t_{r}\ge T$, and $t_{r}\ge 5T$.
Metric values in (a) --- using samples only from $S^{+}$ --- serve as a baseline (all baseline metrics are moderately positive but omitted).
\textbf{Absolute differences} by columns are included in the rest of the table, except that $\widetilde{\rho}$ and $\mathtt{AUC}_{5T}$ values are relative to the baseline of $\rho$ and $\mathtt{AUC}_{T}$, respectively.

Improvements in (b) demonstrate the advantage of incorporating negative affinity samples $S^{-}$.
Results in (c) to (f) show the effectiveness of confidence based sample weighting.
We can see that convex weighting is superior to linear weighting, which in turn exceeds concave weightings.
This implies that emphasizing high confidence observations is helpful.
In addition, we found that $\rho - \widetilde{\rho}<0.03$ except in (a).
This means after normalization for entity types, the predicted affinities still capture the correlation to play durations.
Performance gain with respect to three entity types are consistent, meaning the models tend not to be dominated by a single entity type.
$\mathtt{AUC}_{T}$ in (f) is lower than that in (c) to (d), whereas $\mathtt{AUC}_{5T}$ in (f) exceeds that in (c) to (d).
This suggests linear weighting has better prediction power for $t_{r}\ge T$, whereas convex quadratic weighting achieves better characterization of affinity in longer duration range.
Finally, we found adding more observations further improved the performance in (g).


\subsection{Evaluation per entity}
\label{sec:exp:ent}
Unlike radio stations which have unbounded content, song tracks have bounded lengths (unless they are on repeat).
We investigate how well the model is capturing personalized preferences with respect to each individual entity.
We compute $\rho(e)$ and collect its histogram in each entity type.
Note that not all $e\in E$ have sufficient observations in the test set for a statistically meaningful estimate of $\rho(e)$. 
We focused on the set $\{e|e\in E,\ C_{test}(e)>10,\ p(\rho(e))<0.01\}$, where $C_{test}(\cdot)$ and $p(\cdot)$ denote the count in test set and \textit{p}-value, respectively.
Duplicate $(u,e)$ tuples in the test set have identical affinity prediction, and may inflate $\rho(e)$ due to tie-breaking in ranking the predicted affinities.
Hence only the first occurrence of $(u,e)$ is taken.

\begin{figure}[thb]
  \centering
  \includegraphics[width=\columnwidth]{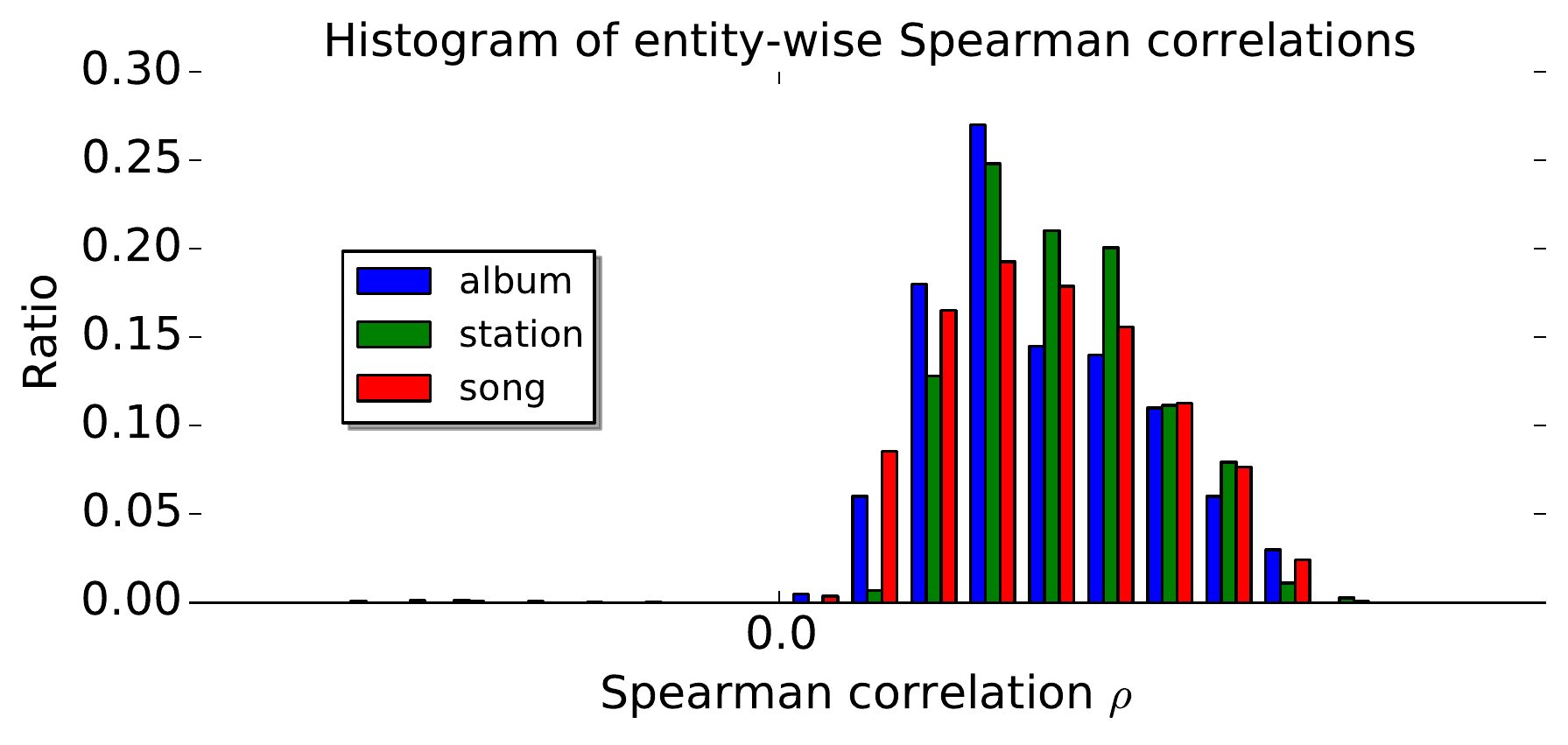}
  \caption{Histograms of entity-wise Spearman correlation $\rho(e)$. \label{fig:hist}}
\end{figure}

As a result, Figure \ref{fig:hist} shows the histograms (absolute values are omitted) computed based on the model in Table \ref{tab:res}, row (g). 
We can see that consistently for three entity types, $\rho(e)$ has a unimodal distribution and $\rho(e)>0$ for almost all $e$.
This lends further support to the effectiveness of the model in capturing personalized affinities to each individual entity.

\subsection{Evaluation on seen \vs unseen interactions}
\label{sec:exp:comp}

Despite causal split of the data, the test set has significant overlap in terms of $(u,e)$ tuples with the training set, due to frequent recurrence of customer usage.
We hence divide the test set by whether the $(u,e)$ tuple has been seen in the training set, and evaluate performances in each part as shown in Table \ref{tab:unseen}.
The first row presents metrics of row (g) in Table \ref{tab:res}.
The rest of the table includes absolute differences.

As expected, there is a significant reduction of performance on the unseen part.
But surprisingly, reduction on the seen part is similar.
For song track and station, performance on the unseen part is even better.
The reason might be heavily biased customer behavior in the seen part.
The system as well as the customers evolve as time goes.
There are a lot less negative affinities in the seen part.
On the system side, SLU errors get fixed over time, so that given the same request, affinity may become positive.
On the customer side, successful requests may be much more likely to reoccur than failed ones.
We conjecture both effects led to a narrower dynamic range of play durations in the seen part, causing a reduction of perceived correlation, as discriminating fine levels of positive affinities is more difficult than separating negative affinities from positive ones. 
This suggests temporal dynamics might be a non-negligible factor in understanding user preference.


\renewcommand{\arraystretch}{0.9}
\begin{table}[!h]
  \caption{Correlation (\textbf{abs. diff.}) on seen \vs unseen $(u,e)$ tuples}
  \label{tab:unseen}
  \centering
  \begin{tabular}{c|ccccc}
    \toprule
    \textbf{Test set} & $\rho$ & $\widetilde{\rho}$ & $\rho_{\mathtt{song}}$ & $\rho_{\mathtt{album}}$ & $\rho_{\mathtt{station}}$ \\
    \midrule
    All & -- & -- & -- & -- & -- \\
    \midrule
    Seen   & $-0.062$  & $-0.065$ & $-0.093$ & $-0.029$ & $-0.086$ \\
    \midrule
    Unseen & $-0.087$  & $-0.043$ & $-0.018$ & $-0.084$ & $+0.039$ \\
    \bottomrule
  \end{tabular}
\end{table}

\section{Conclusion}
\label{sec:con}
The unique characteristics of fully voice-based multimedia services demand innovations in modeling user-entity affinity.
In this work, we have proposed to capture implicit affinities exhibited in play durations.
We used the BPR objective and low-rank matrix factorization to model binarized positive \vs negative affinities.
In the experiment, we found that utilizing positive and negative samples together outperformed using positive samples only; weighting samples with a convex quadratic function yielded the best outcome; and adding more training data further improved the result.
We further proved the model's effectiveness by evaluating on the basis of individual entities, and gained insights about dynamics of system and customer behavior in analyzing performance for seen \vs unseen interactions.

In the future, we would like to cover more entity types including audio books and videos, following a multi-view learning approach \cite{elkahky2015multi}.
We may also incorporate heterogeneous side information such as explicit feedback, similarity of items, \etc \cite{koren2009matrix, he2016vbpr, chen2017learning}.
Moreover, we may extend the affinity matrix to a 3D tensor, following the multi-verse CF method based on tensor factorization \cite{karatzoglou2010multiverse}.
For example, the user's mention text may be added so that the affinity prediction is aware of the user's request.
We would like to further evaluate the effectiveness of affinity prediction in the production environment for the optimization of SLU interpretation and customer experience.

\bibliographystyle{IEEEtran}
\bibliography{user_entity_affinity}

\end{document}